\documentstyle[12pt]{article}          
\makeatletter  

\@addtoreset{equation}{section} \makeatother

\begin{document}

\vspace*{0.5cm}

\newcommand{\hs}{\hspace*{0.5cm}}
\newcommand{\vs}{\vspace*{0.5cm}}
\newcommand{\be}{\begin{equation}}
\newcommand{\ee}{\end{equation}}
\newcommand{\bea}{\begin{eqnarray}}
\newcommand{\eea}{\end{eqnarray}}
\newcommand{\ben}{\begin{enumerate}}
\newcommand{\een}{\end{enumerate}}
\newcommand{\nn}{\nonumber}
\newcommand{\crn}{\nonumber \\}
\newcommand{\non}{\nonumber}
\newcommand{\noi}{\noindent}
\newcommand{\al}{\alpha}
\newcommand{\la}{\lambda}
\newcommand{\bet}{\beta}
\newcommand{\ga}{\gamma}
\newcommand{\va}{\varphi}
\newcommand{\om}{\omega}
\newcommand{\pa}{\partial}
\newcommand{\fr}{\frac}
\newcommand{\bc}{\begin{center}}
\newcommand{\ec}{\end{center}}
\newcommand{\Ga}{\Gamma}
\newcommand{\de}{\delta}
\newcommand{\De}{\Delta}
\newcommand{\ep}{\epsilon}
\newcommand{\varep}{\varepsilon}
\newcommand{\ka}{\kappa}
\newcommand{\La}{\Lambda}
\newcommand{\si}{\sigma}
\newcommand{\Si}{\Sigma}
\newcommand{\ta}{\tau}
\newcommand{\up}{\upsilon}
\newcommand{\Up}{\Upsilon}
\newcommand{\ze}{\zeta}
\newcommand{\ps}{\psi}
\newcommand{\Ps}{\Psi}
\newcommand{\ph}{\phi}
\newcommand{\vph}{\varphi}
\newcommand{\Ph}{\Phi}
\newcommand{\Om}{\Omega}
\def\lappeq{\mathrel{\rlap{\raise.5ex\hbox{$<$}}
{\lower.5ex\hbox{$\sim$}}}}

\bc {\Large \bf $\mbox{U}(1)_Q$ invariance and
$\mbox{SU}(3)_C\otimes \mbox{SU}(3)_L \otimes \mbox{U}(1)_X$\\
 models with $\beta$ arbitrary}

\vspace*{1cm}

{\bf Phung Van Dong$^a$} and {\bf  Hoang Ngoc Long$^b$}\\

\vspace*{0.5cm}

$^a$ {\footnotesize \sl Department of Theoretical Physics, Hanoi
University of Science, Hanoi, Vietnam}\\ $^b$ {\footnotesize \sl
Institute of Physics, VAST, P. O. Box 429, Bo Ho, Hanoi 10000,
Vietnam}

\ec

\begin{abstract}

Using the $\mbox{U}(1)_Q$ invariance, the photon eigenstate and
 matching gauge coupling constants in
 $\mbox{SU}(3)_C\otimes \mbox{SU}(3)_L \otimes \mbox{U}(1)_X$
models with $\beta$ arbitrary are given. The mass matrix of
neutral gauge bosons is exactly diagonalized, and the photon
eigenstate is independent on the symmetry breaking parameters -
VEV's of Higgs scalars. By obtaining the electromagnetic vertex,
the model is embedded naturally into the standard model.

\end{abstract}

PACS number(s): 12.60.Cn, 12.60.Fr, 12.10.Kt, 14.70.Br\\

\noindent

\section{Introduction}

\hs The detection of neutrino oscillation \cite{neuos}
experimentally indicates that neutrinos are massive particles and
that flavour lepton number is not conserved. Since in the standard
model  (SM), neutrinos are massless and  flavour lepton number is
conserved. The neutrino oscillation experiments are a clear sign
that the SM  has to be extended.

A very common alternative to solve some of the problems of the SM
consists of enlarging the group of gauge symmetry, where the
larger group embeds properly the SM. For instance, the
$\mbox{SU}(5)$ grand unification model \cite{vd} can unify the
interactions and predicts the electric charge quantization, while
the group $\mbox{E}_6$ can also unify the interactions and might
explain the masses of the neutrinos \cite{gr}, and etc. \cite{ps}.
Nevertheless, such models cannot explain the generations number
problem.

A very interesting alternative to explain the origin of
generations comes from the cancelation of quiral anomalies
\cite{agg}. In particular, the models with gauge group $G_{331}=
\mbox{SU}(3)_C\otimes \mbox{SU}(3)_L \otimes \mbox{U}(1)_X$, also
called 3-3-1 models \cite{ppf,flt,recent}, arise as a possible
solution to this puzzle, since some of such models require the
three generations in order to cancel chiral anomalies completely.
An additional motivation to study this kind of models comes from
the fact that they can also predict the charge quantization
\cite{prs}.

In the literature about the 3-3-1 models, it is known that the
matching of gauge coupling constants at the $\mbox{SU}(3)_L\otimes
\mbox{U}(1)_X$ breaking is dependent on the constraints among the
VEV's \cite{ppf}. In addition, the independence on the VEV's of
the photon eigenstate and mass has not been explained yet
\cite{Pst}.

In this paper, we have pointed out that the photon eigenstate is
independent on the VEV's; and the matching of gauge coupling
constants is not dependent on VEV's structure.

The paper is organized as follows. In Sec.2 we recall some
features of the 3-3-1 models with $\beta$ arbitrary and study the
mass Lagrangian of the neutral gauge bosons, the photon eigenstate
and mass. Matching gauge coupling constants and diagonalizing the
neutral bosons gauge mass matrix are obtained in Sec.3. Finally,
our conclusions are summarized in the last section.

\section{Photon eigenstate}

\hs The 3-3-1 model with $\beta$ arbitrary \cite{rdm1} has the
electric charge operator in the following form\be Q = T_3+\beta
T_8+X,\label{Q}\ee where $T_3=\la_3/2,T_8=\la_8/2$ are the
$\mbox{SU}(3)_L$ gauge charges, and $X$ is the $\mbox{U}(1)_X$
gauge charge.

Under the gauge symmetry $G_{331}$, the fermion representations
are given in the triplets {\bf 3}, antitriplets {\bf 3$^*$}, and
singlets {\bf 1} (for the right-handed counterparts) of the
$\mbox{SU}(3)_L$ group. In order to cancel anomalies, the same
number of fermion triplets and antitriplets must be present
\cite{rdm1}.

A triplet of the $\mbox{SU}(3)_L$ group is composed from a doublet
{\bf 2} and a singlet {\bf 1} of the $\mbox{SU}(2)_L$ group of the
SM, therefore
it is decomposed as follows \be \left(%
\begin{array}{ccc}
  u, & d, & s \\
\end{array}%
\right)^T=\left(%
\begin{array}{cc}
  u, & d \\
\end{array}%
\right)^T\oplus s, \crn\ee or \be
  (3,X)=(2,X)\oplus(1,X),
 \ee where u, d, and s denote the first,
 the second members of the doublets and of the singlets, respectively.
 In the case of an antitriplet {\bf 3$^*$}, it is also decomposed
 into a antidoublet {\bf 2$^*$} and a singlet of the $\mbox{SU}(2)_L$ group \be
 \left(%
\begin{array}{ccc}
  d,&-u,&s' \\
\end{array}%
\right)^T=\left(%
\begin{array}{cc}
  d,&-u \\
\end{array}%
\right)^T\oplus s', \crn \ee or \be
(3^*,-X)=(2^*,-X)\oplus(1,-X).\ee To find the hyper-charge $Y$ of
the doublets and the singlets, we should use $Y=2(\beta T_8+X)$
which is obtained directly from (\ref{Q}).

The spontaneous symmetry breaking from the $G_{331}$ to the
$G_{SM}$ group of the SM \cite{rdm2} will allow the singlet member
separated from the triplets or antitriplets and get mass. This is
archived by a Higgs scalar triplet $\chi$ with the VEV as follows
\be \langle\chi\rangle^T=\left(0,0,\fr{v_s}{\sqrt{2}}\right).\ee
Then the neutral gauge bosons of the theory get mass from \be
{\cal{L}}^{\chi}_{mass}=(D^H_\mu\langle\chi\rangle)^+
(D^{H\mu}\langle\chi\rangle),\ee where subscripts $H$ denotes
diagonal part of the covariant derivative \be D^H_\mu= \pa_\mu +
igT_3W^3_\mu+igT_8W^8_\mu+ig_X T_9 X_\chi B_\mu. \label{DH} \ee
Here $g$ and $g_X$ are the gauge coupling constants of the
$\mbox{SU}(3)_L$ and $\mbox{U}(1)_X$ groups, respectively.
$X_\chi$ is the $\mbox{U}(1)_X$ charge of the $\chi$ Higgs scalar.
$T_9=diag(1,1,1)/\sqrt{6}$ is chosen such that
$Tr(T_aT_b)=\de_{ab}/2$; $a,b=1,2,...,9$. Substituting the charge
$X_\chi$ from (\ref{Q}) into (\ref{DH}), we get \be D^H_\mu=
\pa_\mu
+igT_3W^3_\mu+igT_8W^8_\mu+i\fr{g_X}{\sqrt{6}}B_\mu\left(Q-T_3-\beta
T_8\right).\label{DH1} \ee The $U(1)_Q$ invariance requires
$Q\langle\chi\rangle=0,$ therefore we get \bea
D^H_\mu\langle\chi\rangle &=& \fr{igv_s}{2\sqrt{2}}
\left(-\fr{2}{\sqrt{3}}W^8_\mu+\fr{2t}{\sqrt{6}}B_\mu\fr{\beta}{\sqrt{3}}\right)\crn
&=&\fr{igv_s}{2\sqrt{2}}\Delta_{3\mu}, \eea where the following
notations are used\bea \Delta_{3\mu} &\equiv&
\left(-\fr{2}{\sqrt{3}}W^8_\mu+\fr{2t}{
\sqrt{6}}B_\mu\fr{\beta}{\sqrt{3}}\right),\\
t &\equiv& g_X/g.\eea\\
Hence\be{\cal{L}}^{\chi}_{mass}=\fr{g^2v_s{^2}}{8}\Delta^2_3,\label{Mchi}\ee
where\[ \Delta^2_3=\Delta_{3\mu}\Delta^{\mu}_3.\]

In the second step of symmetry breaking \cite{agg,ppf,rdm2}, the
$G_{SM}$ group must be decomposed into the
$\mbox{SU}(3)_C\otimes\mbox{U}(1)_Q$ group, two $\eta, \rho$ Higgs
triplets are introduced with following VEV's\bea
\langle\eta\rangle^T &=& \left(\fr{v_u}{\sqrt{2}},0,0\right), \crn
\langle\rho\rangle^T &=& \left(0,\fr{v_d}{\sqrt{2}},0\right).\eea
The neutral gauge bosons also gain mass from two Lagrangians given
by \bea {\cal{L}}^{\eta}_{mass}&=&(D^H_\mu\langle\eta\rangle)^+
(D^{H\mu}\langle\eta\rangle),\\
{\cal{L}}^{\rho}_{mass}&=&(D^H_\mu\langle\rho\rangle)^+
(D^{H\mu}\langle\rho\rangle).\eea Noting that
$Q\langle\eta\rangle=Q\langle\rho\rangle=0$, we get \bea
D^H_\mu\langle\eta\rangle &=& \fr{igv_u}{2\sqrt{2}}
\left[W^3_\mu+\fr{1}{\sqrt{3}}W^8_\mu+\fr{2t}{\sqrt{6}}B_\mu\left(-\fr
1 2 -\fr{\beta}{2\sqrt{3}}\right)\right]\crn
&=&\fr{igv_u}{2\sqrt{2}}\Delta_{1\mu},\\
D^H_\mu\langle\rho\rangle &=& \fr{igv_d}{2\sqrt{2}}
\left[-W^3_\mu+\fr{1}{\sqrt{3}}W^8_\mu+\fr{2t}{\sqrt{6}}B_\mu\left(\fr
1 2 -\fr{\beta}{2\sqrt{3}}\right)\right]\crn
&=&\fr{igv_d}{2\sqrt{2}}\Delta_{2\mu}, \eea here \bea
\Delta_{1\mu}&\equiv&\left[W^3_\mu+\fr{1}{\sqrt{3}}W^8_\mu
+\fr{2t}{\sqrt{6}}B_\mu\left(-\fr
1 2 -\fr{\beta}{2\sqrt{3}}\right)\right],\\
\Delta_{2\mu}&\equiv&\left[-W^3_\mu+\fr{1}{\sqrt{3}}W^8_\mu
+\fr{2t}{\sqrt{6}}B_\mu\left(\fr
1 2 -\fr{\beta}{2\sqrt{3}}\right)\right].\eea Hence \bea
{\cal{L}}^{\eta}_{mass}&=&\fr{g^2v^2_u}{8}\Delta^2_1,\crn
{\cal{L}}^{\rho}_{mass}&=&\fr{g^2v^2_d}{8}\Delta^2_2.\label{Mer}\eea
Finally, the mass Lagrangian of the neutral gauge bosons is given
by \bea
{\cal{L}}^{NCC}_{mass}&=&{\cal{L}}^{\eta}_{mass}+{\cal{L}}^{
\rho}_{mass}+{\cal{L}}^{\chi}_{mass},\crn
&=&\fr{g^2}{8}\left(v^2_u\Delta^2_1+v^2_d\Delta^2_2+v^2_s\Delta^2_3\right)\label{Lm}.
\eea

In general, any 3-3-1 model need to have three Higgs triplets for
breaking the $G_{331}$ group into the
$\mbox{SU}(3)_C\otimes\mbox{U}(1)_Q$ group. However, some 3-3-1
models need less than three Higgs \cite{dgp}. For our purpose in
obtaining the mass matrix of the neutral gauge bosons, this
corresponds to vanishing of $v_u$ or $v_d$. In the case with more
than three Higgs triplets, one just makes the following
appropriate replaces \bea v^2_u&\rightarrow&
v^2_u+v^2_{u1}+v^2_{u2}+...,\crn v^2_d&\rightarrow&
v^2_d+v^2_{d1}+v^2_{d2}+...,\crn v^2_s&\rightarrow&
v^2_s+v^2_{s1}+v^2_{s2}+...,\label{DS}\eea where
$v_{ui},v_{dj},v_{sk}$ are the VEV's of the additional Higgs
triplets. They belong to the up, down, and singlet members,
respectively. This also remains correctly for the cases, if a
Higgs triplet has two neutral members with the non-zero VEV's
\cite{dgp}, and for models with Higgs antitriplets.

In some models, for example the minimal 3-3-1 model \cite{ppf}, to
give mass to all leptons, we have to introduce a Higgs sextet. Let
us denote the Higgs sextet by $\Gamma_{ij}$. Then, the mass
Lagrangian will get an addition \be
{\cal{L}}^{\Gamma}_{mass}=(D^H_\mu\langle\Gamma\rangle_{ij})^+
(D^{H\mu}\langle\Gamma\rangle_{ij}),\label{LS}\ee where \bea
D^H_\mu\langle\Gamma\rangle_{ij}&=&ig \left[(W^3_\mu T_3+W^8_\mu
T_8)^k_i \langle\Gamma\rangle_{kj}+ (W^3_\mu T_3+W^8_\mu T_8)^k_j
\langle\Gamma\rangle_{ki}\right]+\fr{ig_X}{\sqrt{6}}X_{\Gamma}B_\mu
\langle\Gamma\rangle_{ij}\crn &=&ig \left[\left(W^3_\mu
T_3+W^8_\mu T_8+\fr{t}{\sqrt{6}}(Q-T_3-\beta T_8)B_\mu\right)^k_i
\langle\Gamma\rangle_{kj}\right]\crn &+& ig\left[ \left(W^3_\mu
T_3+W^8_\mu T_8+\fr{t}{\sqrt{6}}(Q-T_3-\beta T_8)B_\mu\right)^k_j
\langle\Gamma\rangle_{ki}\right]\crn &=& \fr{ig}{2}\left(%
\begin{array}{ccc}
  2\langle\Gamma\rangle_{11}\Delta_{1\mu} & \langle\Gamma\rangle_{12}(
  \Delta_{1\mu}+\Delta_{2\mu}) & \langle\Gamma\rangle_{13}(\Delta_{1\mu}+\Delta_{3\mu}) \\
  \langle\Gamma\rangle_{12}(\Delta_{1\mu}+\Delta_{2\mu}) & 2\langle
  \Gamma\rangle_{22}\Delta_{2\mu} & \langle\Gamma\rangle_{23}(\Delta_{2\mu}+\Delta_{3\mu}) \\
  \langle\Gamma\rangle_{13}(\Delta_{1\mu}+\Delta_{3\mu}) & \langle\Gamma
  \rangle_{23}(\Delta_{2\mu}+\Delta_{3\mu}) & 2\langle\Gamma\rangle_{33}\Delta_{3\mu} \\
\end{array}%
\right).\eea It is easy to verify that \be
\Delta_{1\mu}+\Delta_{2\mu}+\Delta_{3\mu}=0.\ee Finally, the mass
term for neutral gauge bosons from sextet is given by \be
{\cal{L}}^{\Gamma}_{mass}=\fr{g^2}{2}\{[2\langle\Gamma\rangle^2_{11}+
\langle\Gamma\rangle^2_{23}]\Delta^2_1+[2\langle\Gamma\rangle^2_{22}+
\langle\Gamma\rangle^2_{13}]\Delta^2_2+[2\langle\Gamma\rangle^2_{33}+
\langle\Gamma\rangle^2_{12}]\Delta^2_3\}.\label{Lcc} \ee Note
that, only neutral members in sextet can have non-zero VEV's. From
(\ref{Lcc}) we see that the general form of mass Lagrangian
(\ref{Lm}) is not changed by adding ${\cal{L}}^{\Gamma}_{mass}.$

In order to generate the fermion masses, the Higgs bosons should
lie in either the triplet, antitriplet, sextet, or singlet
representation of $\mbox{SU}(3)_L$ \cite{rdm2}. In later case, the
singlet must be neutral, and it does not give mass to gauge
bosons. So, we conclude that for any 3-3-1 model, the mass matrix
of the neutral gauge bosons always has the form given in
(\ref{Lm}).

The mass Lagrangian (\ref{Lm}) can be rewritten \be {\cal
L}_{mass} = \fr 1 2 V^T M^2 V, \ee where $V^T=(W^3,W^8,B)$, and
\be
M^2=\fr 1 4 g^2\left(%
\begin{array}{ccc}
  m_{11} & m_{12} & m_{13} \\
  m_{12} & m_{22} & m_{23} \\
  m_{13} & m_{23} & m_{33} \\
\end{array}%
\right), \ee with \bea m_{11}&=&v^2_u+v^2_d,\crn
m_{12}&=&\fr{1}{\sqrt{3}}\left(v^2_u-v^2_d\right),\crn
m_{13}&=&\fr{t}{\sqrt{6}}\left[v^2_u\left(-1-\fr{\beta}{
\sqrt{3}}\right)-v^2_d\left(1-\fr{\beta}{\sqrt{3}}\right)\right],\crn
m_{22}&=&\fr{1}{3}\left(v^2_u+v^2_d+4v^2_s\right),\crn
m_{23}&=&\fr{t}{3\sqrt{2}}\left[v^2_u\left(-1-\fr{\beta}{
\sqrt{3}}\right)+v^2_d\left(1-\fr{\beta}{\sqrt{3}}
\right)-v^2_s\fr{4\beta}{\sqrt{3}}\right],\crn
m_{33}&=&\fr{t^2}{6}\left[v^2_u\left(-1-\fr{\beta}{\sqrt{3}}
\right)^2+v^2_d\left(1-\fr{\beta}{\sqrt{3}}\right)^2+v^2_s\left(
\fr{2\beta}{\sqrt{3}}\right)^2\right].\eea

It can be checked that the matrix $M^2$ has a {\it non degenerate}
zero eigenvalue for any breaking parameters in any  3-3-1 model.
Therefore, the zero eigenvalue is identified with the photon mass,
$M^2_\ga=0$.

The eigenstate with the zero
eigenvalue can be obtained directly from the following equation \be M^2 \left(%
\begin{array}{c}
  A_{\ga1} \\
  A_{\ga2} \\
  A_{\ga3} \\
\end{array}%
\right)=0. \ee
We get then \be A_\ga=\left(%
\begin{array}{c}
   t \\
  \beta t \\
  \sqrt{6} \\
\end{array}%
\right) \fr{1}{\sqrt{6+(1+\beta^2)t^2}}. \ee So, the physical
photon field $A_\mu$ is given by\be
A_\mu=\fr{t}{\sqrt{6+(1+\beta^2)t^2}}W^3_\mu+\fr{\beta
t}{\sqrt{6+(1+\beta^2)t^2}}W^8_\mu+
\fr{\sqrt{6}}{\sqrt{6+(1+\beta^2)t^2}}B_\mu.\label{ga}\ee

For any 3-3-1 model, we see that the photon eigenstate and mass
are not dependent on the VEV's ($v_u,v_d,v_s$). These are a
natural consequence of the $\mbox{U}(1)_Q$ invariance - the
conservation of the electric charge.

\section{Matching gauge coupling constants}

\hs To embed the 3-3-1 model into the SM, we will work with
electromagnetic vertex. Let us denote two remain massive fields by
$Z^1_\mu,Z^2_\mu$. We change basis by unitary matrix \be
(A_\mu,Z^1_\mu,Z^2_\mu)=(W^3_\mu,W^8_\mu,B_\mu)U
\label{change},\ee where $U$ has the following form
\be U=\left(%
\begin{array}{ccc}
  \fr{t}{\sqrt{6+(1+\beta^2)t^2}} & U_{12} & U_{13} \\
 \fr{\beta
t}{\sqrt{6+(1+\beta^2)t^2}} & U_{22} & U_{23} \\
 \fr{\sqrt{6}}{\sqrt{6+(1+\beta^2)t^2}} & U_{32} & U_{33} \\
\end{array}%
\right).\label{unitn}\ee Here elements in the second and third
columns are not necessary to determine.  From (\ref{change}) and
(\ref{unitn}), we get \bea
W^3_\mu&=&\fr{t}{\sqrt{6+(1+\beta^2)t^2}}A_\mu+U_{12}Z^1_\mu+U_{13}Z^2_\mu,
\crn W^8_\mu&=&\fr{\beta
t}{\sqrt{6+(1+\beta^2)t^2}}A_\mu+U_{22}Z^1_\mu+U_{23}Z^2_\mu, \crn
B_\mu&=& \fr{\sqrt{6}}{\sqrt{6+(1+\beta^2)t^2}} A_\mu+U_{32}
Z^1_\mu+U_{33}Z^2_\mu \label{b}. \eea

The interactions among the gauge bosons and fermions are given
by\bea {\cal L}_F &=&
\bar{R}i\ga^\mu\left(\pa_\mu+i\fr{g_X}{\sqrt{6}}XB_\mu\right)R
\crn &+& \bar{L}i\ga^\mu\left(\pa_\mu+igW^a_\mu
\fr{\la_a}{2}+i\fr{g_X}{\sqrt{6}}XB_\mu\right)L,  \label{Lf}\eea
where $R$ represents any right-handed singlet and $L$-any
left-handed triplet or antitriplet. Substituting $W^3, W^8, B$
from (\ref{b}) into (\ref{Lf}), for $R=e_R$ with $X_{e_R}=-1$ and
$L=(\nu_{eL},e_L,E_{L})^T$ with $X_L=-1/2-\beta/2\sqrt{3}$, we get
\bea {\cal L}^{int}_{\bar{e}e\ga} &=& -\bar{e}_R i \ga^\mu
\left[\fr{ig_X}{\sqrt{6}}\fr{\sqrt{6}}{\sqrt{6+(1+\beta^2)t^2}}\right]A_\mu
e_R \crn &+&\bar{e}_L i \ga^\mu
\left[-\fr{ig}{2}\fr{t}{\sqrt{6+(1+\beta^2)t^2}}+\fr{ig}{2\sqrt{3}}\fr{\beta
t}{\sqrt{6+(1+\beta^2)t^2}}\right.\crn &-&\left.
\fr{ig_X}{\sqrt{6}}\left(\fr 1 2
+\fr{\beta}{2\sqrt{3}}\right)\fr{\sqrt{6}}{\sqrt{6+(1+\beta^2)t^2}}
\right]A_\mu e_L \crn
&=&\fr{g_X}{\sqrt{6+(1+\beta^2)t^2}}\bar{e}\ga^\mu e A_\mu.
\label{eeg} \eea

The coefficient of the $\bar{e}e\ga$ vertex in (\ref{eeg}) is
identified with the electromagnetic coupling constant \be
\fr{g_X}{\sqrt{6+(1+\beta^2)t^2}} \equiv e. \label{eggx} \ee In
the SM, we have \be \fr{g_{2}g_Y}{\sqrt{g^2_2+g^2_Y}}=e,
\label{eg'g} \ee where $g_2$, $g_Y$ are coupling constants of
$SU(2)_L$ and $U(1)_Y$ gauge group, respectively.  Using
continuation of gauge coupling constant of $SU(3)_L$ group at the
spontaneous symmetry breaking point,\be g=g_2\equiv g,\ee from
(\ref{eggx}) and (\ref{eg'g}), we get \be
\fr{1}{g^2_Y}=\fr{\beta^2}{g^2}+\fr{6}{g^2_X}. \label{mtm}\ee From
(\ref{eggx}), we obtain \be
\fr{t}{\sqrt{6+(1+\beta^2)t^2}}=\fr{e}{g}. \ee As in the SM, we
put \be \fr{t}{\sqrt{6+(1+\beta^2)t^2}}=\sin\theta_W. \ee The
(\ref{mtm}) yields \be
t=\fr{\sqrt{6}t_W}{\sqrt{1-\beta^2t^2_W}}.\ee Hence, the photon
eigenstate can be rewritten in the form \be A_\mu=s_W W^3_\mu +
c_W\left(\beta t_W  W^8_\mu + \sqrt{1-\beta^2 t^2_W}B_\mu
\right).\label{ga1} \ee
 From orthogonal condition of the photon
eigenstate to two remain gauge vectors, we can write \bea
Z_\mu&=&c_W W^3_\mu - s_W\left(\beta t_W  W^8_\mu +
\sqrt{1-\beta^2 t^2_W}B_\mu
\right),\label{Z} \\
Z'_\mu&=&\sqrt{1-\beta^2 t^2_W}W^8_\mu -\beta t_WB_\mu.\label{Z'}
\eea Therefore, in this basis, the mass matrix $ M^2 \rightarrow
M^{2'}$
has the form as follows \be M^{2'}=\left(%
\begin{array}{ccc}
  0 & 0 & 0 \\
  0 & M^2_Z & M^2_{ZZ'} \\
  0 & M^2_{ZZ'} & M^2_{Z'} \\
\end{array}%
\right), \ee where \bea M^2_Z &=&
\fr{g^2}{4c^2_W}(v^2_u+v^2_d),\crn
M^2_{ZZ'}&=&\fr{g^2}{4\sqrt{3}c^2_W\sqrt{1-(1+\beta^2)s^2_W}}\{[(\sqrt{3}
\beta-1)s^2_W+1]v^2_u+[(\sqrt{3}\beta+1)s^2_W-1]v^2_d\},\crn
 M^2_{Z'}
 &=&\fr{g^2}{12(1-\beta^2 t^2_W)}\left[(1+\sqrt{3}\beta t^2_W)^2v^2_u
 +(1-\sqrt{3}\beta t^2_W)^2v^2_d+4v^2_s\right].
\eea
 The matrix, $M^{2'},$ gives mixtures between $Z_\mu$ and
$Z'_\mu$,
 by rotating an angle $\phi$ in the plane
 $(Z_\mu,Z'_\mu)\rightarrow(Z^1_\mu,Z^2_\mu)$, the mass eigenvectors are \bea
 Z^1_\mu &=& Z_\mu\cos\phi - Z'_\mu\sin\phi,     \crn
 Z^2_\mu &=& Z_\mu\sin\phi + Z'_\mu\cos\phi,\eea where $\phi$ is defined by \be
 \tan ^2\phi=\fr{M^2_Z-M^2_{Z^1}}{M^2_{Z^2}-M^2_Z}, \ee and the
 physical mass eigenvalues:\bea M^2_{Z^1}&=&\fr 1 2 \left[M^2_Z+M^2_{Z'}
 -\sqrt{(M^2_Z-M^2_{Z'})^2+4(M^2_{ZZ'})^2}\right],\\
 M^2_{Z^2}&=&\fr 1 2
 \left[M^2_Z+M^2_{Z'}+\sqrt{(M^2_Z-M^2_{Z'})^2+4(M^2_{ZZ'})^2}\right].\eea

From the mixing mass matrix of $Z$ and $Z'$ we see that $\phi=0$
if  $v_s\gg v_u, v_d$ or \be
 v^2_u=\fr{1-(\sqrt{3}\beta+1)s^2_W}{1+(\sqrt{3}\beta-1)s^2_W}v^2_d \nn\ee
Here $A, Z^1$ correspond to the neutral
 gauge bosons the SM $(\ga, Z)$, and $Z^2$ is a new neutral gauge boson.

To finish this section, we note that the matching condition of the
coupling constants (\ref{eggx}) at the $SU(3)_L\otimes U(1)_X$
breaking is very obvious as the matching in the SM. It is not
dependent on the constraint $v_s\gg v_u, v_d$ as in the literature
\cite{ppf}. After the matching, we rewrote the photon field with
the coefficients in the Weinberg mixing angle, and then taking
exact diagonalization of the mass matrix for the neutral gauge
bosons.
\section{Conclusion}

\hs In this paper, the photon eigenstate and the matching of
coupling constants in  3-3-1 models are obtained in general form
containing Higgs triplets, antitriplets as well as sextet. We
emphasized that the matching of coupling constants is not
dependent on condition that vacuum expectation value of Higgs
boson of the first step of breaking symmetry must be much larger
than those of the second step namely $\langle s \rangle \gg
\langle u\rangle, \langle d\rangle$.

This technique can be extended for electroweak models which are
based on the larger gauge groups such as
$\mbox{SU}(3)_C\otimes\mbox{SU}(4)_L\otimes\mbox{U}(1)_X$~\cite{su4}.

This work was supported in part by National Council for
Natural Sciences of Vietnam contract No: KT - 41064.\\[0.3cm]

\end{document}